\documentclass{article}
\usepackage{arxiv}

\usepackage[utf8]{inputenc} 
\usepackage[T1]{fontenc}    
\usepackage{url}            
\usepackage{booktabs}      
\usepackage{amsfonts}       
\usepackage{nicefrac}    
\usepackage{microtype}      
\usepackage{lipsum}
\usepackage{float}
\usepackage{amsmath}
\usepackage{graphicx}
\usepackage[colorinlistoftodos]{todonotes}
\usepackage[colorlinks=true, allcolors=blue]{hyperref}
\usepackage{wrapfig}

\newcommand{\lengthPBS}{210~$\mu$m }
\newcommand{\SplitDb}{14.37~dB }
\newcommand{\lengthDC}{140~$\mu$m }
\newcommand{\PBS}{polarisation beamsplitter}
\newcommand{\PCF}{PCF}  
\newcommand{\PCFs}{PCFs}

\author{
  Andrea Bertoncini \\
  Biological and Environmental Science and Engineering Division\\ King Abdullah University of Science and Technology (KAUST) \\ Thuwal, Saudi Arabia\\
   \And
 Carlo Liberale \\
  Biological and Environmental Science and Engineering Division\\ King Abdullah University of Science and Technology (KAUST) \\ Thuwal, Saudi Arabia\\
  \texttt{carlo.liberale@kaust.edu.sa} \\
}
\title{3D printed waveguides based on Photonic Crystal Fiber designs for complex fiber-end photonic devices}
\begin{document}
\maketitle

\begin{abstract}

	Optical waveguide segments based on geometrically unbound photonic crystal fibers (PCF) designs could be exploited as building blocks to realize miniaturized complex devices which implement advanced photonic operations. 
	Here, we show how to fabricate optical waveguide segments with PCF designs by direct high-resolution 3D printing and how the combination of these segments can realise complex photonic devices. We demonstrate the unprecedented precision and flexibility of our method by fabricating the first-ever fiber polarising beam splitter based on PCFs. 
	The device was directly printed in one step on the end-face of a standard single-mode fiber and was 210~$\mu$m-long, offering broadband operation in the optical telecommunications C-band.
	Our approach harnesses the potential of high-resolution 3D printing and of PCF designs paving the way for the development of novel miniaturised complex photonic systems, which will positively impact and advance optical telecommunications, sensor technology, and biomedical devices.	
	\end{abstract}

\section{Introduction}
Photonic Crystal Fibers (PCFs), also known as microstructured optical fibers or Holey Fibers, are single-material optical fibers in which an array of microscopic longitudinal hollow channels enables light guidance \cite{Russell:2003gv,knight2003photonic}. The design of the geometry of the longitudinal hollow channels in PCFs is a powerful handle to control and tune the fiber waveguide parameters, such as optical mode size and shape, modal dispersion, birefringence, and nonlinearity. 
With the development of \PCFs, unprecedented fine control of the fiber waveguide parameters across a wider range has become achievable, opening up unique possibilities like supercontinuum generation \cite{dudley2009ten}, fiber chromatic dispersion engineering \cite{saitoh2003chromatic} and ultra-high birefringence \cite{ortigosa2000highly}. Furthermore, \PCFs \space are unique in allowing the creation of hollow-core fibers \cite{cregan1999single}, which have important applications such as fiber propagation with ultra-low nonlinearity or novel gas and optofluidic sensors \cite{cubillas2013photonic}. 

Optical waveguides based on \PCF \space designs could be exploited on the small scale as building blocks to create on-fiber complex miniaturized devices which implement advanced photonic operations including --- but not limited to --- mode conversion, Y-splitting, and  polarization splitting. For such devices, the accurate and geometrically unbound manufacture of the designed \PCF \space transverse hole patterns is of paramount importance.
Additionally, precise control of the longitudinal variation of the \PCF \space geometry allows to create elements such as ultra-short adiabatic tapers or periodic structures, which will pave the way for the development of novel miniaturised photonic devices.
However, current \PCF \space fabrication methods have important limitations in manufacturing \PCF \space segments with the needed characteristics and for their union to create complex miniaturized photonic systems.
\PCFs \space are primarily fabricated by drawing a cylindrical "preform" of cm-scale diameter \cite{Russell:2006ih}. The preform has a cross-sectional geometry that corresponds to a scaled-up version of the desired final sub-mm-scale geometry of the fiber. 
Current methods to create the preform, however, grant only limited freedom in the design of the preform \cite{peng20193d}. Additionally, during the drawing process, the preform geometry is generally not preserved due to material viscosity, gravity, and surface tension effects \cite{PCF_fab}. Therefore,
obtaining the desired \PCF \space cross-sectional structure is not a straightforward process, and can be especially difficult. Specific hole geometries are even impossible to realize \cite{Ringcore_fabricated}.  
The 3D printing of cm-scale \PCF \space preforms has been recently proposed as a means to increase freedom of design, but the perturbing effects of drawing still present a major limiting factor that prevents the accurate realisation of arbitrary \PCF \space designs \cite{peng20193d,Cook:2015bm,van2019design,talataisong2018mid}. 
Lastly, $\mu$m-scale control of the length of \PCF \space segments and of their longitudinal tapering, which are needed to create miniaturised photonic systems, is very difficult with standard preform-based methods.

Here, we show the use of high-resolution 3D printing \cite{kawata2001finer,malinauskas2016ultrafast,jonuvsauskas2019femtosecond} for the in-situ single-step fabrication of stacked ultra-short \PCF-like segments with different geometries to create all-fiber integrated devices that perform complex optical operations in sub-mm lengths.
Our approach entirely avoids the drawing process that introduces so many limitations and drawbacks, and grants unprecedented design flexibility and precision in the control of the transverse and longitudinal \PCF \space geometry.
We begin by demonstrating that our proposed approach can precisely replicate the hole array geometry for virtually any class of manufactured \PCF \space designs described in the literature.
We then establish the manifold advantages of our approach for the miniaturisation of complex optical devices by fabricating a 210~$\mu$m-long broadband, all-fiber, and integrated Polarisation BeamSplitter (PBS), which is the first \PCF \space PBS ever realised and described in the literature.

\section{Results}
\label{La carrellata}

\begin{figure*}
	\begin{center}
		\centering
		\includegraphics[width=1\linewidth]{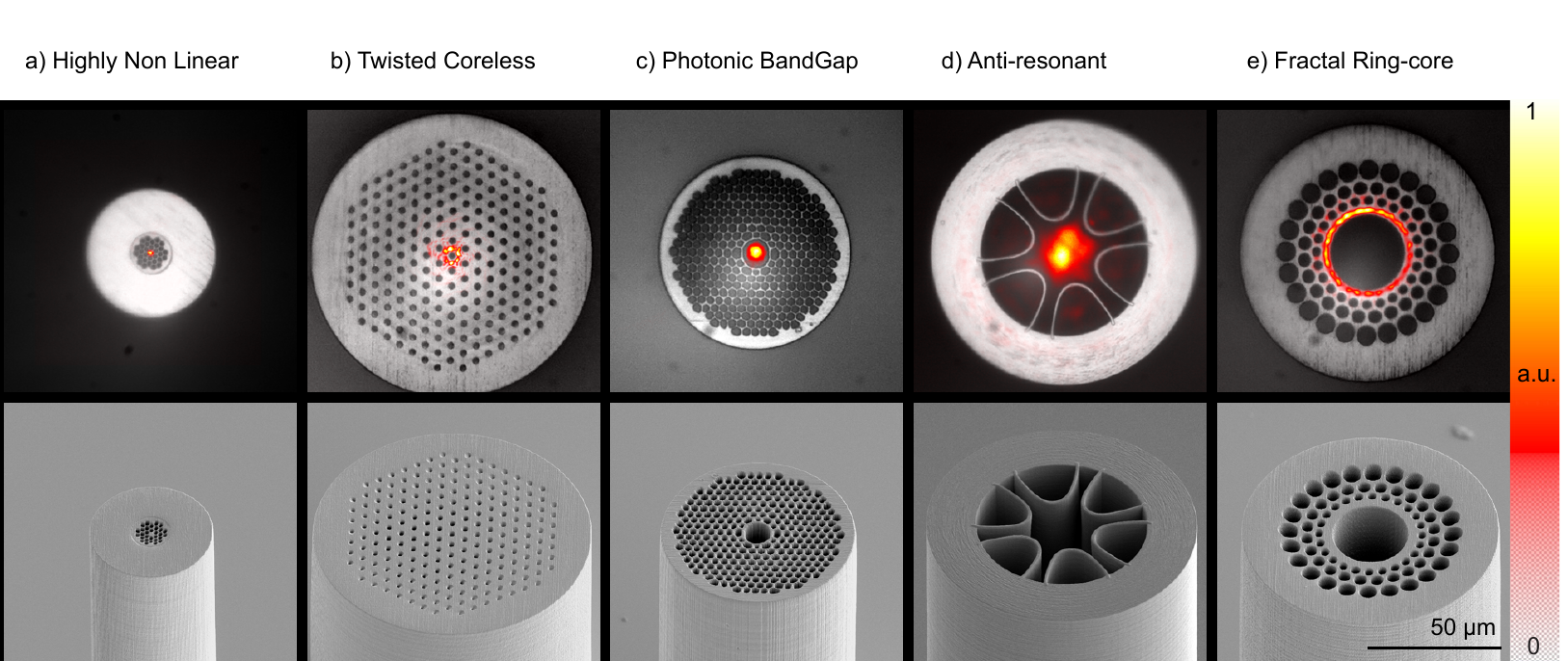}
		\caption{\textbf{The optical guidance of different classes of 3D printed PCF designs demonstrate the viability of the method}: a) Highly Non Linear \PCF; b) Helically twisted core-less \PCF \space (the arrow indicates the twist direction); c) Photonic Bandgap Hollow-core \PCF; d) Anti-resonant Hollow-core \PCF; e) Fractal Ring-core \PCF. The top row shows the output optical mode overlaid with the optical microscopy image of the 3D printed waveguide end-face; the bottom row shows the corresponding Scanning Electron Microscopy (SEM) images. The wavelength for the optical mode is 1060~nm for all of the structures, except for b), which is 640~nm.}
		\label{fig:carrellata}
	\end{center}
\end{figure*}

\textbf{3D printing of different classes of Photonic Crystal Fiber designs.}
In this section, we present Scanning Electron Microscope (SEM) images and the optical guidance of 3D printed \PCF-like segments with various solid or hollow-core geometries to demonstrate the accuracy and flexibility of the proposed method (see \nameref{Methods} for fabrication details). For demonstration, we replicated an assortment of known \PCF \space designs based on radically different guiding mechanisms, core shapes and sizes.
Figure \ref{fig:carrellata}a shows a 3D printed segment with the typical design of a \textit{Highly Non Linear} (HNL) \PCF \space \cite{leon2004supercontinuum}, with a core diameter of 2~$\mu$m, an air-filling fraction --- defined as the ratio of the air-hole diameter \textit{d} to the lattice spacing $\Lambda$ --- equal to 0.75, and a Mode Field Diameter (MFD) of 1.8~$\mu$m. This type of \PCF \space is typically characterized by a small core (few $\mu$m in diameter), with hexagonal holes and a high air-filling fraction. The guiding mechanism in HNL \PCFs \space is based on Modified Total Internal Reflection (MTIR) \cite{Russell:2003gv} --- analogous to that of a standard single-mode fiber --- whereby the pattern of holes surrounding the central core acts as an effective cladding with a reduced refractive index \cite{firstPCFref}.
We directly printed the HNL \PCF \space segment on the end-face of a single-mode fiber with a 6~$\mu$m MFD. The large modal mismatch between the optical fiber and the 3D printed segment was easily compensated for by including a 70~$\mu$m long \PCF-like adiabatic taper in the 3D printed structure, similar to that described in \cite{bertoncini2017fiber}, which resulted in a 1.7~dB insertion loss (see Supplementary information section  \ref{suppl:1mm}).
Figure \ref{fig:carrellata}b shows a 3D printed \textit{helically twisted core-less} \PCF \space segment with the same geometrical pattern proposed by \cite{CorelessPCF}. Here, the twist of the fiber around its axis induces guidance next to the central hole, which is not twisted. 
Increasing the twist rate makes the guided mode more confined and less sensitive to perturbations.
Thanks to the high resolution granted by our 3D printing method, we could easily achieve a very high twist rate of 10*pi~[rad/mm] (corresponding to a twist period of 200~$\mu$m), which is higher than any twist rate previously reported in the literature. As expected for this type of \PCF \space design, we obtained a well-defined hollow mode confined to the first ring around the central non-twisted hole, with an MFD of 6.57~$\mu$m.
Next, we fabricated hollow-core \PCF-like structures based on two different guiding mechanisms: \textit{Photonic Bandgap Hollow-core} fibers and \textit{Hollow-core anti-resonant} fibers \cite{PCF_HC_review_Benanid}.
In Photonic Bandgap Hollow-core fibers, optical confinement is provided by the Photonic Bandgap (PBG) mechanism, in which the periodic array of holes in the cladding acts as a photonic crystal that prohibits the propagation of light, which is then trapped in the hollow core.
We fabricated a PBG \PCF-like segment with a geometry that replicated a commercially available fiber (HC-1060-02, NTK Photonics), and the final structure showed the expected light guidance in the central hole satisfactorily (Fig. \ref{fig:carrellata}c), with an MFD of 8.2~$\mu$m.
In Hollow-core anti-resonant fibers (HC-ARFs), the light is confined through a combination of inhibited coupling between the core and cladding modes and anti-resonant reflection at the air-fiber-material interfaces. The hollow-core region is defined by anti-resonant elements with negative curvature.
We 3D printed a HC-ARF geometry that replicated a more recent design \cite{van2019design}, where the anti-resonant elements are semi-elliptical. Semi-elliptical elements are typically problematic to manufacture using traditional preform drawing-based methods \cite{chaudhuri2016low,van2019design}. However, here we show that semi-elliptical structures can be easily and accurately reproduced by 3D printing. Figure \ref{fig:carrellata}d shows an accurate reproduction of the structure and the expected guidance in the central hollow core, with an MFD of 12.1~$\mu$m.
Note that hollow-core PBG and HC-ARF fibers rely on guiding mechanisms that are very sensitive to the geometric precision of the structures. The optical guidance that we achieved here intrinsically demonstrates that our 3D printed \PCF \space structures are geometrically accurate.
Finally, Fig. \ref{fig:carrellata}e shows a \textit{Fractal Ring-core} \PCF-like segment \cite{Ringcore_fabricated}, which supports a well-defined annular mode through an MTIR guidance mechanism.
These types of structures are attracting increasing interest because they have been recently shown to support modes that carry Orbital Angular Momentum \cite{Ringcore_fabricated}.

Traditional \PCFs \space manufactured by drawing require a final fiber cleavage step, which can eventually distort the final fiber structure or create non-flat output surfaces. Remarkably, 3D printing of \PCF-like waveguides is not affected by this issue because the 3D printing process allows to directly produce flat perpendicular or angled output surfaces. 

We measured the propagation losses of the 3D printed \PCF-like waveguides by fabricating segments of different lengths, up to 350~$\mu$m. 
For a solid-core \PCF \space design with a core size of 12~$\mu$m and $d$/$\Lambda$ = 0.4, we found an attenuation of 0.44~dB/mm at 1070~nm and of 0.79~dB/mm at 1550~nm. For comparison, a pure silica fiber with the same \PCF \space geometry (ESM 12B, Thorlabs) has an attenuation of about 8~dB/km. The propagation losses of the 3D printed \PCF \space segment closely match the extinction coefficient for the bulk polymerised photoresist that are 0.43~dB/mm at 1070~nm and 0.78~dB/mm at 1550~nm \cite{schmid2019optical}, which is then the dominant loss contribution (see also the \nameref{sec:Conclusion} section and supplementary section \ref{losses_supp}).
For the 3D printed waveguide with a PBG Hollow-core \PCF \space design presented in Fig. \ref{fig:carrellata}c, we found an attenuation of 0.3~dB/mm  at 1070~nm. This attenuation, while being lower than the intrinsic photo-polymerised material losses, is not as low as expected for propagation in a hollow core. 
This could be explained with the fact that a dominant factor in hollow-core \PCF \space losses is the surface roughness of the core wall.
While pure silica hollow-core \PCFs \space have typically a sub-nanometer root mean square (RMS) roughness value, in our case the 3D printing layer-by-layer fabrication introduced a larger RMS roughness of about 30~nm (estimated from SEM images). This roughness value is consistent to what measured by other groups using the same 3D printing technology and material \cite{dietrich2018situ}.

\textbf{Design and fabrication of an all-fiber integrated \PCF \space \PBS}.
Here, we present an on-fiber ultrashort \PBS  \space (PBS) based on a dual-core \PCF \space design to demonstrate the multiple strengths of our 3D printing approach.
Several sub-mm dual-core \PCF \space PBS designs have been proposed over recent years \cite{Jiang:2014io,PCFsplit1,PCFsplit3,PCFsplit4}); however, the limitations of current \PCF \space fabrication methods have prevented their successful manufacture. Indeed, the dual-core geometries that have been proposed in the literature to date have all been generally asymmetric, with the inclusion of holes of different sizes and shapes, all factors that add significant complexity to the design of the preform.
Moreover, these \PCF \space PBS designs have a sub-mm length that requires precise control to a sub-$\mu m$ level to create the desired output polarisation split. These combined factors make it difficult to handle and cut segments to the required length from a long fiber that has been drawn.
Furthermore, on-fiber integration of the \PCF \space \PBS \space requires rigid coupling to a standard fiber, e.g., by fusion splicing. This coupling also requires a small but critical lateral offset of a few micrometres in order to directly couple just one of the two cores of the dual-core structure. This integration step is also significantly challenging with \PCFs \space manufactured by fiber drawing.

\begin{figure*}[h]
	\centering
	\includegraphics[width=1\textwidth]{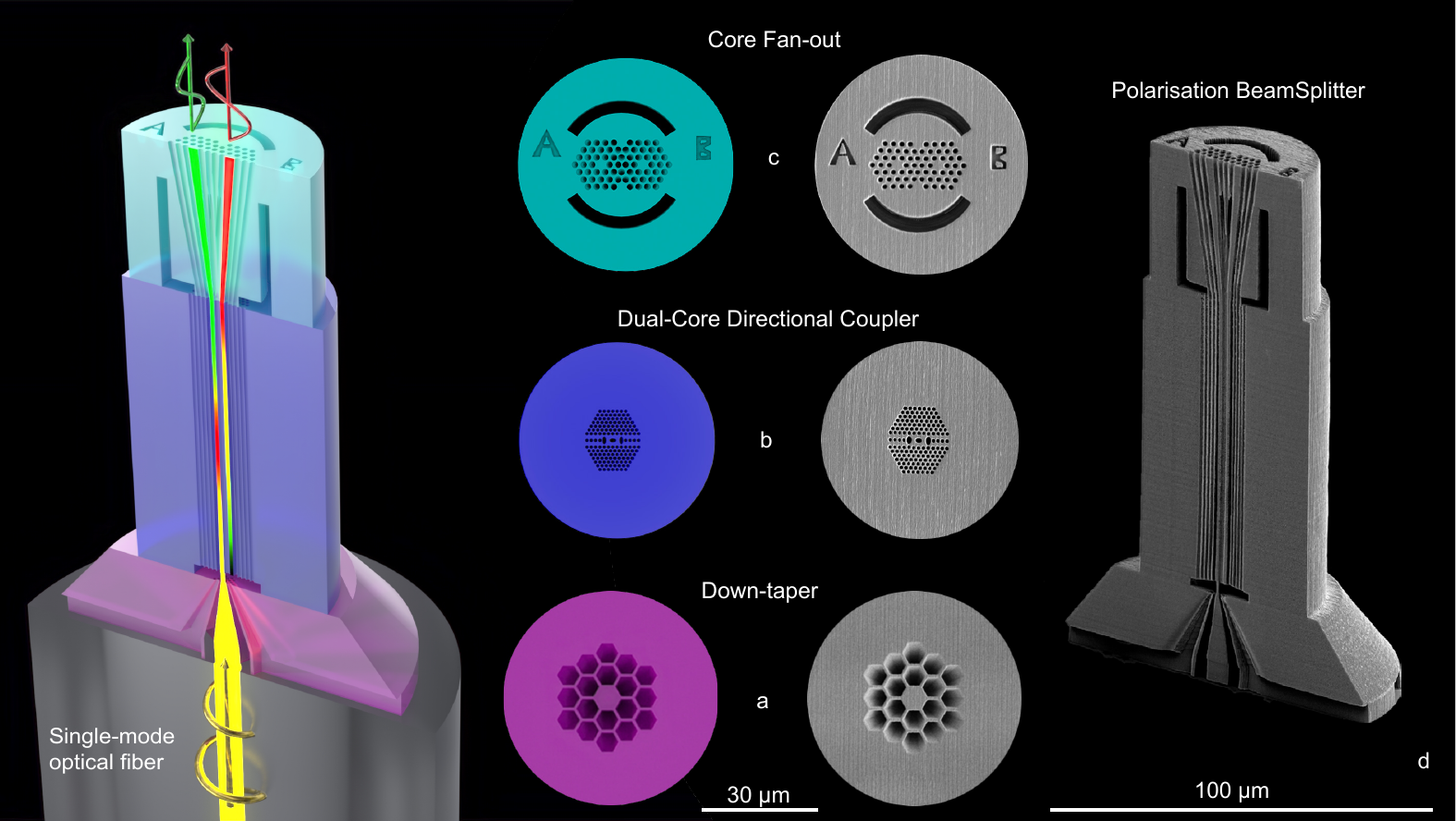}
	\caption{\textbf{The first fabrication of a \PCF \space polarisation beamsplitter enabled by the 3D printing of \PCF \space designs.}
		\label{fig:PBS_Graphical_Abstract}
		Left side: a rendered illustration of the \PCF \space PBS, where three different \PCF \space segments are integrated on the end-face of a single-mode fiber:
		a down-taper section (magenta); a dual-core Directional Coupler (DC) section (blue); a core fan-out section (cyan). A beam with an arbitrary  polarisation (yellow beam) is split into its horizontal (red beam) and vertical (green beam) polarisation components.
		Right side: SEM image of the cross-section of the complete 3D printed \PCF \space PBS.
		Center: side-by-side display of the three different rendered cross-sections and corresponding SEM images from 3D printed structures.
	}
	\label{fig:PBS_Graphical_Abstract}
\end{figure*}

Figure \ref{fig:PBS_Graphical_Abstract}b shows the design for a dual-core \PCF \space PBS theoretically proposed by Jiang et al. \cite{Jiang:2014io}, which we have chosen to implement here, with necessary modifications, as a demonstration. This design has many exceptional merits, such as featuring a very large bandwidth that includes the telecommunication C-band and an ultrashort length.
Note that the ultrashort length is achieved thanks to the ultrahigh birefringence enabled by the \PCF \space design.
The dual-core \PCF \space segment acts as a directional coupler (DC), and is characterised by a coupling length (CL) for each polarisation, defined as the waveguide length for which there is a complete transfer of power from one core to the other. 
In particular, the coupling lengths are given by:

\begin{equation}
	\label{CL_eq}
	CL_i = \frac{\lambda}{2 * (n_i^e - n_i^o)}
\end{equation}

where $\lambda$ is the wavelength, $n_i^e$ and $n_i^o$ are the effective indexes for the even and odd mode of the dual-core waveguide, respectively, and \textit{i} = x,y is either of the two orthogonal polarizations. 
Because of the birefringence introduced by this \PCF \space design, the two CLs are different, which allows the structure to act as a PBS for a proper tuning of its design parameters and at specific lengths \cite{Jiang:2014io}. The length of the dual-core DC \PCF \space structure must be simultaneously an odd integer multiple of the CL for one polarisation and an even integer multiple of the CL for the other polarisation \cite{Jiang:2014io}. The shortest possible polarisation splitting dual-core DC is obtained when the length of the structure is equal to the CL for one polarisation and twice the CL for the other polarisation, thus giving a CL ratio (CLR) of 2.

Efficient integration of the dual-core DC \PCF \space structure on a standard single-mode optical fiber requires the addition of other elements. By leveraging one of the strengths of the 3D printing approach, we have embedded the dual-core DC \PCF \space structure in a more complex photonic structure composed of three sequential waveguiding segments (Fig. \ref{fig:PBS_Graphical_Abstract}): a \PCF-like tapered coupler (down-taper), the dual-core DC birefringent \PCF \space structure, and a final fan-out section
that increases the spatial separation of the two cores.
The down-taper (input cross-section in Fig. \ref{fig:PBS_Graphical_Abstract}a, simulations in supplementary section \ref{taper_supp}) allows for efficient and alignment-free coupling of a 6~$\mu$m MFD single-mode fiber (1060XP, Thorlabs) to one of the two cores of the birefringent dual-core DC \PCF \space segment (Fig. \ref{fig:PBS_Graphical_Abstract}b).
The two cores of the latter are non-circular, relatively small (1~$\mu$m along the minor axis), and positioned close to each other (2.4~$\mu$m apart) to maximise the core inter-coupling and obtain the shortest possible CLs. 
The final segment is a fan-out structure with a \PCF \space design, which rapidly spatially separates the two cores up to a 10~$\mu$m distance, to facilitate optical measurements of the PBS outputs (output cross-section in Fig. \ref{fig:PBS_Graphical_Abstract}c).
This segment also provides a simple solution for coupling to other optical fibers or for integration into optical chip components by allowing a modal re-shape of the two orthogonally polarised output beams; in our case, an adiabatic transformation from an asymmetric 1x2~$\mu$m mode to a 3~$\mu$m diameter round mode (see supplementary section \ref{fan-out}).
We set the working spectral range for the PBS to be centered at 1550~nm, thus covering the optical communications C-band.

A design of an optimal (CLR=2) dual-core DC \PCF \space segment based uniquely on the calculation of the modal effective indexes, and the use of Eq. \ref{CL_eq}, cannot account for several aspects of the entire real-world design-to-fabrication process, such as the discretised geometry in the 3D printing system and possible anisotropic shrinkage of the structures during post-exposure development \cite{gissibl2016two}. 
These effects could make the fabricated \PCF-like structure geometrically deviate slightly from the wanted \PCF \space design. Additionally, it is difficult to simulate the role of the transition from the  dual-core DC \PCF \space segment to the down-taper and the fan-out sections (See Supplementary information sec. \ref{Suppl:CLsim_notworking}).
For this reason, we defined the final design of the complete \PCF \space PBS structure using an iterative approach that involved modal analysis, fabrication and optical measurements; this iterative approach was enabled by the fast turnover time achievable by 3D printing. We used the modal analysis to provide reliable guidelines on how the CLs change with size variations of different parts of the structure's geometry.
In each step of this iterative optimisation process, we selected a different geometrical parameter of the dual-core \PCF \space segment to be varied, based on its effect on the CLs for the two orthogonal polarisations, and hence on the CLR, as indicated by the numerical calculations with modal analysis. Then, we fabricated an array of different PBS structures on a glass coverslip, where each structure had a different value for the selected geometrical parameter.
We generated the initial guess for the dual-core \PCF \space geometry from modal analysis of a geometry very similar to the one presented in \cite{Jiang:2014io}, while accounting for the refractive index of the used photopolymer (1.532 at 1550~nm, see \nameref{Methods} and reference \cite{schmid2019optical}).
The structures in each array were individually coupled with a focused free-space beam, and their output sections were imaged onto an InGaAs infrared camera to extract, for each polarisation, the ratio between the powers carried by the two cores (see \nameref{Methods}).

\begin{figure}[t]
	\centering
	\includegraphics[width=1\linewidth]{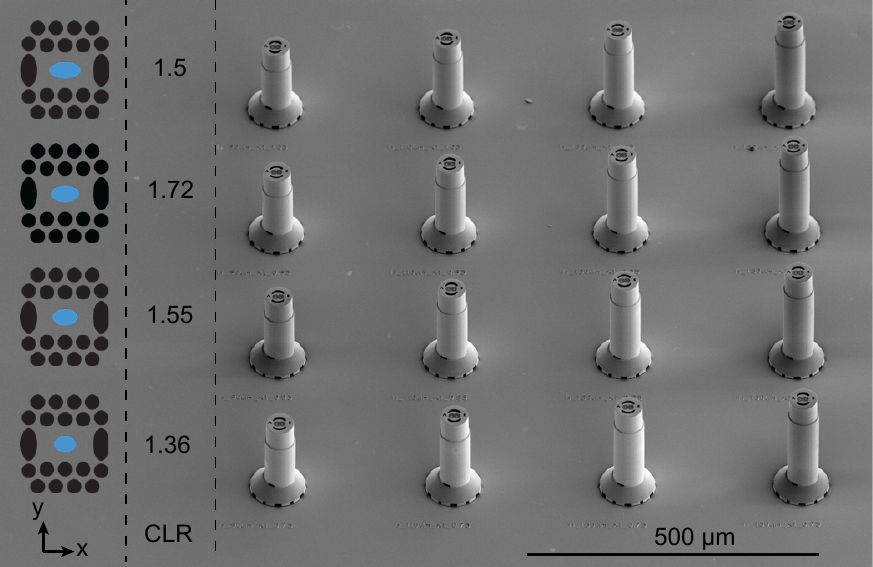}
	\caption{\textbf{3D printed array for iterative fast optimisation of the \PCF \space PBS structure.} 
		Scanning electron micrograph of an array of 3D printed structures used in the CLR optimisation process. Here, we tested \PCF \space PBS structures with four central hole ellipticities: for each ellipticity, corresponding to a row in the array, we printed four different lengths of the dual-core DC \PCF \space section to retrieve its CLR. The fabrication time for this sample is $\sim$ 6 hours. In the geometry tab on the left, the modified geometrical element is highlighted in blue and only the important central holes are displayed.}
	\label{fig:optimization_example}
\end{figure}

\begin{figure}[t]
	\begin{center}
		\centering
		\includegraphics[width=1\linewidth]{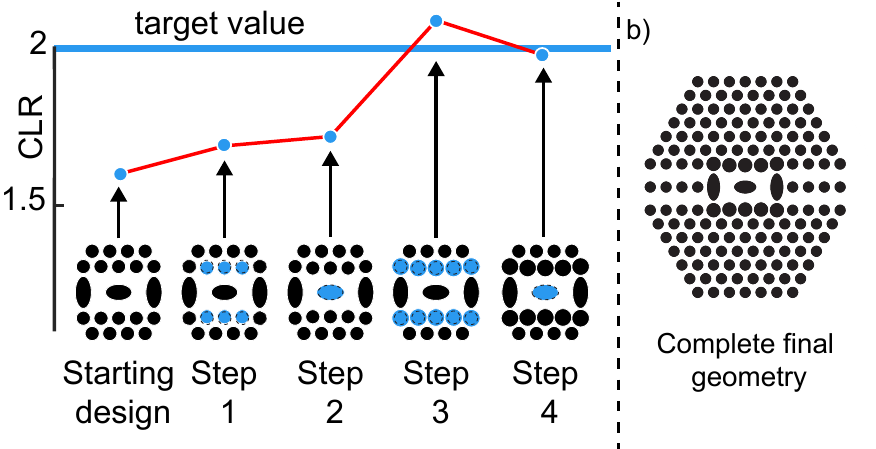}
		\caption{\textbf{Experimental optimisation of the \PCF \space PBS geometry}. a) Progression of the best CLR in each step of the optimisation process towards CLR $\,\to\,$~2. The geometrical elements that have been  changed in each iteration step are highlighted in blue (only the important central holes are displayed);  b) Complete cross-sectional \PCF \space geometry of the optimised \PCF \space PBS design.}
		\label{fig:OptimizationProcess}
	\end{center}
\end{figure}

\begin{figure*}
	\begin{center}
		\centering
		\includegraphics[width=1\linewidth]{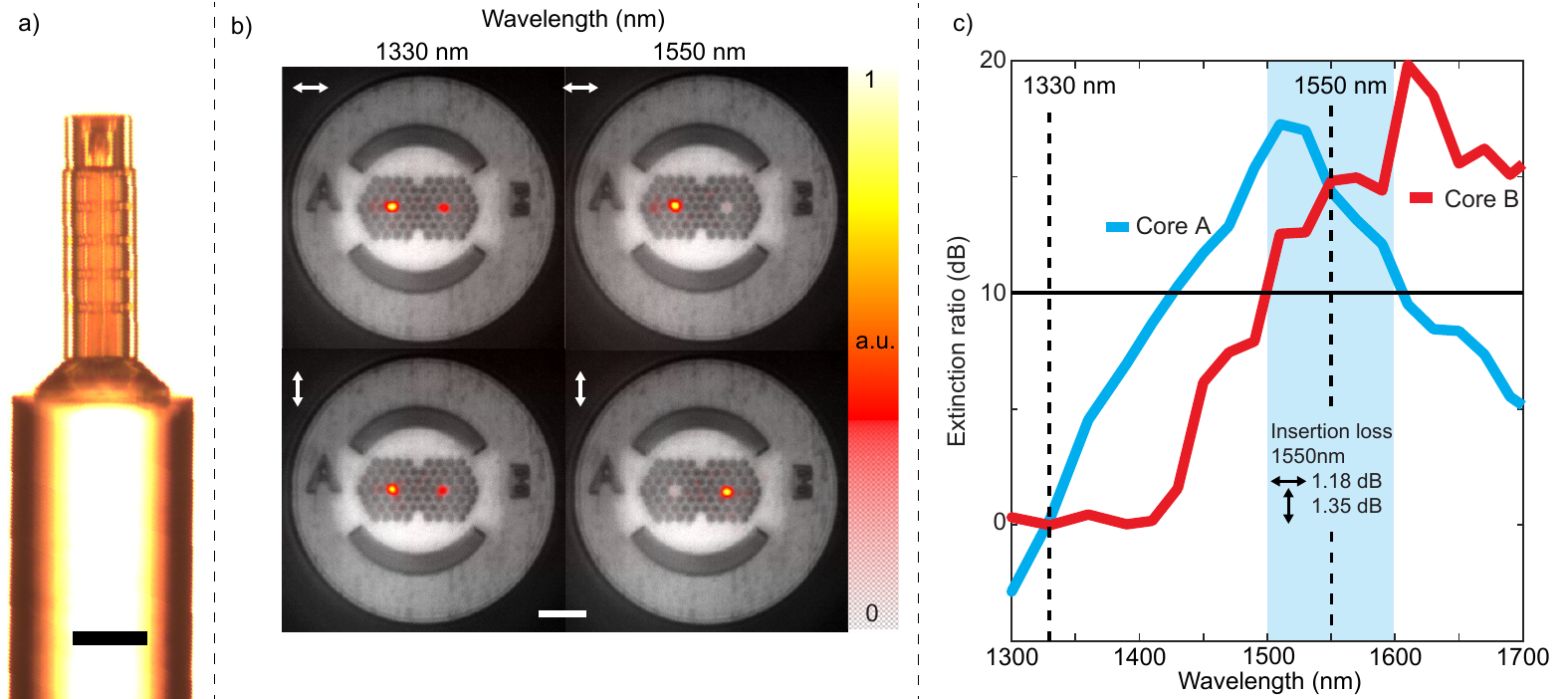}
		\caption{\textbf{Experimental polarisation splitting performance of the \PCF \space PBS 3D printed on a single-mode fiber.} a) Side-view optical image of the \PCF \space PBS (scale bar 50~$\mu$m). b) Output facet of the \PCF \space PBS overlaid with images of the output beams collected for each polarisation (indicated by the white double arrow in each sub-figure) with an InGaAs camera at different wavelengths. At 1330~nm, the polarisations are mixed, whereas at 1550~nm the polarisations are well segregated in different cores (scale bar 10~$\mu$m). c) Extinction ratios on the two cores of the \PCF \space PBS, which simultaneously exceed 10~dB on a 100~nm bandwidth centered at around 1550~nm (highlighted area).}
		\label{fig:finalPBS}
	\end{center}
\end{figure*}

Figure \ref{fig:optimization_example} shows an example of an optimisation iteration step. 
Here, we exploited the dependence of the CLR on the ellipticity of the central hole in the dual-core DC \PCF \space segment.
We printed an array of structures with four different ellipticities, and for each ellipticity, we printed a further three structures with different lengths, increasing from left to right in Fig. \ref{fig:optimization_example}, for a total of 16 structures in the array.
By fitting the variation in the ratio of powers carried by the two cores at different lengths, we could extrapolate the two CLs, hence giving the CLR for each different ellipticity (see Supplementary information  Fig. \ref{fig:CLtrends_suppl}).
We used the structure with the geometry that gave a CLR closest to~2 as a starting point for the next round of the iteration, where we changed a different geometrical parameter using the guidelines provided by the simulations (see supplementary section \ref{Suppl:optimization}). From the new fabricated array  we identified a new geometry that achieved a CLR even closer to~2. 
The complete optimisation process involved four steps and the variation of three geometrical parameters (Fig. \ref{fig:OptimizationProcess}a and supplementary section \ref{Suppl:optimization}), concluding with the optimised design shown in Fig. \ref{fig:OptimizationProcess}b, which gave a satisfactory CLR of~1.97.
The optimised design has a \lengthDC~long dual-core DC \PCF \space segment and a total length of 210~$\mu$m for the complete structure. We directly printed this finalised PCF PBS structure on the end-face of a single-mode fiber (Fig. \ref{fig:finalPBS}a) and characterised its broadband polarisation splitting performance.
This \PCF \space PBS 3D printed on fiber had an extinction ratio of more than 10~dB over a bandwidth of 100~nm and  centered around 1550~nm (Fig. \ref{fig:finalPBS}c). Both cores had an extinction ratio above 12.6~dB in the fiber optics communication C-band (1530-1565~nm). At 1550~nm, we achieved a minimum extinction ratio of \SplitDb (Fig. \ref{fig:finalPBS}b-c).
The insertion loss at 1550~nm was 1.18~dB for the  horizontal polarization and 1.35~dB for the vertical polarization. These insertion losses could be further improved by using a longer down-taper section, to make it adiabatic according to the length-scale criterion \cite{love1991tapered}.
The bandwidth of our \PCF \space PBS was very broad, but not as broad as that predicted by Jiang et al. \cite{Jiang:2014io} (i.e., 150~nm at 10~dB extinction ratio). We can attribute this difference to three main factors: 1) accurate modal analysis of the dual-core DC \PCF \space geometry shows the existence of higher order modes that were neglected in ref. \cite{Jiang:2014io}, but whose contribution degrades the PBS extinction ratio; 2) the final CLR of the 3D printed structure is not exactly equal to~2; 3) some residual scattering from the photopolymerised material may lead to inter-core cross-talk.

\section{Discussion}
\label{sec:Conclusion}
In this work, we first demonstrated the successful direct 3D printing and optical guidance of a selection of optical waveguides with \PCF-like designs that rely on different guiding mechanisms. By successfully fabricating these \PCF \space designs, we verified that our method can achieve the fabrication precision and optical quality required for obtaining the final desired cross-sectional \PCF \space geometry considerably faster than current \PCF \space fabrication methods. 
We then demonstrated that our method is capable of fabricating \PCF-like waveguides with geometries that were previously impossible to manufacture because of their complexity. Specifically, we succeeded in fabricating the first ever  \PCF \space PBS. 
This \PCF \space PBS is the first example of miniaturised complex structures made of stacked segments with \PCF \space designs, presenting fast longitudinal tapers and precisely controlled lateral offsets.
Finally, through the realisation of the \PCF \space PBS, we showed how direct 3D printing of \PCF-like waveguides allows for a comprehensive optimisation process that is significantly faster than current \PCF \space fabrication methods based on the drawing of a preform.
Besides demonstrating the strengths of our approach, the  \PCF \space PBS that we fabricated is significant in itself, as miniaturisation and fibre integration of polarisation splitting devices are highly desirable features, especially in optical communication systems. 
Polarising beam splitters 3D printed on optical fibers have been already reported in the literature \cite{Hahn:18,PBS_inverse}, however they are based on diffraction mechanisms and a further integration in a fiber optical system could be complicated by their intrinsic free-space output.
In addition, we anticipate that a similar dual-core \PCF \space structure such as ours could be optimised to perform wavelength demultiplexing or all-optical switching \cite{allopticalswitchdcpcf,Vieweg:12}.
Based on current high-resolution 3D printing technology, the maximum length that can be achieved for a \PCF-like waveguide is in the order of a few mm because of limitations in the fabrication speed and in the configuration of 3D printing machines.
However, we foresee that advances in multiphoton lithography fabrication performance \cite{hahn20193,Saha105} will soon allow for the fabrication of longer segments and at faster speeds. 
3D printing fabrication also opens up the possibility to fabricate the bulk parts of the structures that are not used in light propagation (e.g., the outer part of the waveguide cladding) as a wireframe mesh. With this approach, lighter and faster to fabricate robust structures are achievable, potentially leading to the design of unique opto-mechanical properties \cite{frenzel2017three}. 
Such wireframe structuring is not currently achievable with traditional drawing-based methods. 
The current propagation losses for 3D printed solid-core \PCF-like waveguides are relatively high, and are contributed mainly by the extinction coefficient of the polymerised photoresist \cite{schmid2019optical}, which is significantly higher than that of standard fiber optic materials such as fused silica.
We expect that future improvement in multi-photon polymerisable materials will lead to more favourable propagation losses. Additionally, an approach described recently for high-resolution 3D printing of glass-ceramics could allow the use of less lossy materials \cite{glass_ceramics}, which could also provide better mechanical and thermal properties to the printed \PCF \space segments than what offered by polymers.
The propagation losses of 3D printed Hollow-core \PCF \space designs are also relatively high, in this case mainly because of the intrinsic roughness of longitudinal surfaces, which is two orders of magnitude higher than typical values for drawn glass \PCFs. 
This roughness is determined by the chosen slicing step-size that, while allowing for a reasonable fabrication time, was nevertheless not optimal for reducing propagation losses. 
As fabrication speeds and methods improve in the future, smaller slicing steps will become more viable, leading to smoother surfaces and lower propagation losses.
Nevertheless, even if the current propagation losses of the 3D printed waveguides based on \PCF \space designs are a little too high for long propagation distances, they are still suitably low enough to achieve unique and well-performing miniaturised photonic devices.
We expect that our approach will open up new possibilities to enhance optical fiber end-faces with miniaturised hybrid complex photonic systems based on segments having \PCF \space designs, as well as their easy combination with other 3D-printable refractive, reflective, diffractive and metamaterial-based elements \cite{malinauskas2010femtosecond,liberale2010micro,bertoncini2018polarization,gissibl2016two,Hahn:18}.
These structures may find application in Orbital Angular Momentum, optical tweezers, and quantum technologies.
New, more sophisticated fiber-end probes for biomedical applications may also emerge.
We also foresee the development of novel fiber end-face sensors that use 3D printed hollow-core \PCF \space designs for bioanalytics and optofluidics \cite{de2019prospects}. These applications could benefit from new photoresists with low autofluorescence that are being developed.
The inclusion of metals and liquids in high-resolution 3D printed structures has already been demonstrated \cite{gansel2009gold,toulouse2018alignment}; this technology could be combined with our method to create multi-material hybrid \PCF-like structures \space \cite{hybridPCFreview}.
We also expect that optical and fiber-optic engineers could benefit from the unprecedented possibilities offered by the freedom of design of \PCF \space geometries in several ways: a) the easier fabrication of previously difficult-to-produce \PCF \space geometries could unlock new designs, including not yet proposed designs that were hitherto considered impossible to fabricate; b) several properties (e.g., mode shape, mode size, etc.) of special \PCF \space designs could be experimentally tested without concern for long turnaround times to achieve the desired fiber geometries fabricated.
We also predict that our technology could be applied in the development of twisted optical fibers. In addition to the very high twist rates achievable, a finely controlled transverse and/or axial modulation of the twist rate, as is easily achieved by 3D printing, could lead to new optical effects \cite{Beravat2018}.
Finally, our 3D printing approach to create optical waveguides that exploit the unique properties of PCF-designs, could easily integrate/complement other recently proposed methods that share the same printing technology, for creating and coupling optical waveguides and photonic chips \cite{dietrich2018situ,Photonicwirebonding,landowski2017direct,gordillo2019plug}. 

We anticipate our method will unleash the creativity of \PCF \space designers and enable a new generation of miniaturised on-fiber photonic structures for enhancement of the fiber end-face, positively impacting and advancing optical telecommunications, sensor technology, and biomedical devices.

\section{Materials and methods} 
\label{Methods}

\textbf{Fabrication.}
3D printing through two-photon lithography offers sub-$\mu$m resolution \cite{gan2013three}, three-dimensional design freedom, and has been recently exploited in several fields, including micro-optics \cite{malinauskas2010femtosecond,liberale2010micro,bertoncini2018polarization,gissibl2016two,dietrich2018situ}.
In two-photon lithography, a focused near-IR femtosecond laser beam induces the highly localized polymerisation of a photopolymer \cite{sun2004two}.
All the structures presented in this work were 3D printed by a commercially available two-photon lithography system (Photonic Professional GT, Nanoscribe).
The photopolymer used here was the proprietary IP-Dip (Nanoscribe GmbH), which is the one that provides the highest fabrication resolution among those available from Nanoscribe. This photoresist is mainly composed of pentaerythritol triacrylate \cite{IpDip_is_PETA}, and its absorption spectrum can be found in \cite{schmid2019optical}.
We used this resist with a 63x 1.4~NA microscope objective (Zeiss), and in dip-in \cite{dip-in} lithography configuration --- in which the microscope objective is directly dipped into the photoresist. In our printing configuration, the polymerized voxel has an ellipsoidal shape, with a typical size of $\sim$~0.3~$\mu$m~*~1~$\mu$m.
The writing laser is a near-infrared femtosecond fiber laser with a pulse duration of $\sim$ 100~fs, a 780~nm wavelength, and a 80~MHz repetition rate. It uses galvanometric mirrors for beam steering in the system, which allowed a high linear writing speed up to 100~mm/s. The 3D printing was executed layer by layer, with the transverse (\textit{x-y}) scanning performed by the galvo system while the axial (\textit{z}) movement was carried out by a piezo actuator.
The distance between the different exposed lines is usually referred to as "hatching" in the case of the \textit{x-y} plane and as "slicing" for the \textit{z} axis. We used a 0.3~$\mu$m slicing distance, a 0.2~$\mu$m hatching distance, a scan speed of 10~mm/s and a laser power of 13.5~mW. Under these settings, the total fabrication time of the complete structure printed on fiber, which was \lengthPBS long, was around 25 minutes.

Following completion of the 3D printing, the structures were developed in the mr-Dev 600 developer.
To ensure the complete development of the very high aspect ratio hollow channels of the \PCF-like waveguides --- \lengthDC long and 0.7~$\mu$m in diameter in the case of the dual-core DC \PCF \space segment --- we adopted a multi-step strategy. We began with a 5-minute development step to remove the bulk of the unpolymerised photoresist. Then we proceeded with two 20 minute development steps to remove any remaining unpolymerised photoresist from the hollow channels. We then immersed the sample in isopropanol for 25 minutes to remove any remaining developer, and allowed the sample to air-dry. We used fluorescence confocal laser scanning microscopy to asses if the hollow channels were completely developed  (see Supplementary information section \ref{suppl:confoca}).
We fabricated the structures either on glass slides using a standard Nanoscribe substrate holder, or directly on the end-face of single-mode optical fibers (Thorlabs 1060XP) using a custom-designed holder. To guarantee optical fiber alignment and stability during 3D printing, we inserted the fiber in a ferrule, and then terminated and connectorised the fiber. Using this approach, the fiber is more stable compared to using a v-groove based fiber holder. 

For direct 3D printing of the PBG \PCF-like waveguide shown in Fig. \ref{La carrellata}c, we precisely replicated the geometry of the commercially available fiber HC-1060-02 (NTK Photonics), starting with the SEM image of the output face as given in the manufacturer data-sheet. We processed the SEM image to obtain a 2D binary mask that we used directly as the desired transverse geometry in the 3D printer slicing software. Finally, we extruded this adopted 2D design to the desired length and 3D printed it.\\

\textbf{Optical characterisation.}
We assessed the performance of the structures that were 3D printed on a planar glass substrate, e.g., those shown in Fig. \ref{fig:optimization_example}. The structures were optically coupled by focussing the collimated output of a tunable laser beam (SuperK COMPACT supercontinuum laser with SuperK SELECT tunable filter, NTK Photonics) with a 20x 0.4~NA microscope objective (Nikon). 
The polarisation of the input beam was defined by using a linear polariser and a half-wave plate. The output beams from the \PCF-like waveguides were collected with a 40x 0.6~NA microscope objective (Nikon) and imaged on a InGaAs camera (Xeva 320, Xenics) with a 500~mm tube lens, after passing through a linear polariser to select either of the two orthogonal polarisations.  
We collected the images of the \PCF \space PBS output beams with the same integration time for the two orthogonal positions of the output linear polariser. We then processed them in MATLAB to obtain the output power for each core by spatially integrating the image on the respective core areas.
We calculated the extinction ratio for each core from $ER_{A}=10log{10}(P_{Ax}/P_{Ay})$ for core A and $ER_{B}=10log{10}(P_{By}/P_{Bx})$ for core B, where $P_{ix}$ and $P_{iy}$ are the output powers on core \textit{i} (with \textit{i}=A,B) for the x and y polarisation, respectively. 

To assess the optical performance of the \PCF \space structures 3D printed directly on the end-face of single-mode fibers, we used a FC/PC fiber connector to directly couple the optical fiber to the fiber delivery unit of the supercontinuum laser, and we controlled the input polarisation with a fiber polarisation controller (FPC030, Thorlabs).
See Supplementary information \ref{suppl:CLR_calc} for details on the calculations of the coupling length ratios.

\renewcommand\thefigure{\thesection.\arabic{figure}}
\setcounter{figure}{0}

\section*{Acknowledgements}

We gratefully acknowledge financial support from King Abdullah University of Science and Technology (KAUST) through baseline funding BAS/1/1064-01-01.\\

\section*{}

See Supplement 1 for supporting content.
\footnotesize

\bibliographystyle{ieeetr}
\bibliography{revised}

\normalsize
\onecolumn

\pagebreak
\pagenumbering{arabic}

\appendix
\raggedbottom
\renewcommand{\thesubsection}{S\arabic{subsection}}
\renewcommand{\thesection}{S\arabic{section}}
\renewcommand{\thefigure}{S\arabic{figure}}
\renewcommand{\thetable}{S\arabic{table}}
\renewcommand{\theequation}{S\arabic{equation}}
\setcounter{section}{0}

\section*{3D printed waveguides based on Photonic Crystal Fiber designs for complex fiber-end photonic devices: supplemental document}

This document provides supplementary information to “3D printed waveguides based on Photonic Crystal Fiber designs for complex fiber-end photonic devices” Optica volume, first page
(year), http://dx.doi.org/10.1364/optica.0.000000. We provide further details on: the fabrication of
a 1mm long highly non linear PCF-like waveguide, the estimation of propagation losses, the simulations of the PCF PBS (down-taper segment, fan-out segment, and dual-core segment), details on the calculation of the coupling length ratio and on the optimisation procedure of the PCF PBS. Finally, we show the confocal fluorescence microscopy characterization of one PBS PCF structure.

\subsection{Fabrication of a 1~mm-long Highly Non Linear \PCF-like segment}
\label{suppl:1mm}

Figure \ref{suppl:1mm} presents an example of one 1~mm-long 3D-printed \PCF-like segment. 
The truncated-cone structure at the base of the fiber is a down-taper with a \PCF-like design that adiabatically converts the 6~$\mu$m MFD LP01 mode of the standard single mode fiber to a 1~$\mu$m MFD. The hole pattern of the down-taper (Fig. \ref{fig:PBS_Graphical_Abstract}a) is a typical hole pattern found in \textit{Highly Non Linear} \PCFs, with a high air-filling fraction to achieve a small core diameter with high confinement. The tapering angle is constant and is calculated to ensure adiabatic shrinking of the fiber fundamental mode. To test the performance of the down-taper segment alone, we printed it on the end-face of a single mode fiber (1060XP, Thorlabs) that was butt-coupled to a commercial Highly Non Linear \PCF \space (NL-1.5-670-02, Blaze Photonics), which had a core diameter of 1.5~$\mu$m and a 1~$\mu$m MFD. Power measurements gave an insertion loss as low as 1.7~dB at 1060~nm,when using standard mechanical positioning actuators during the butt-coupling alignment. However, when the same fibers were butt-coupled without the 3D printed taper, we instead measured a 6.5~dB insertion loss due to the high modal dimension mismatch.

\begin{figure}[H]
	\centering
	\includegraphics[width=0.7\linewidth]{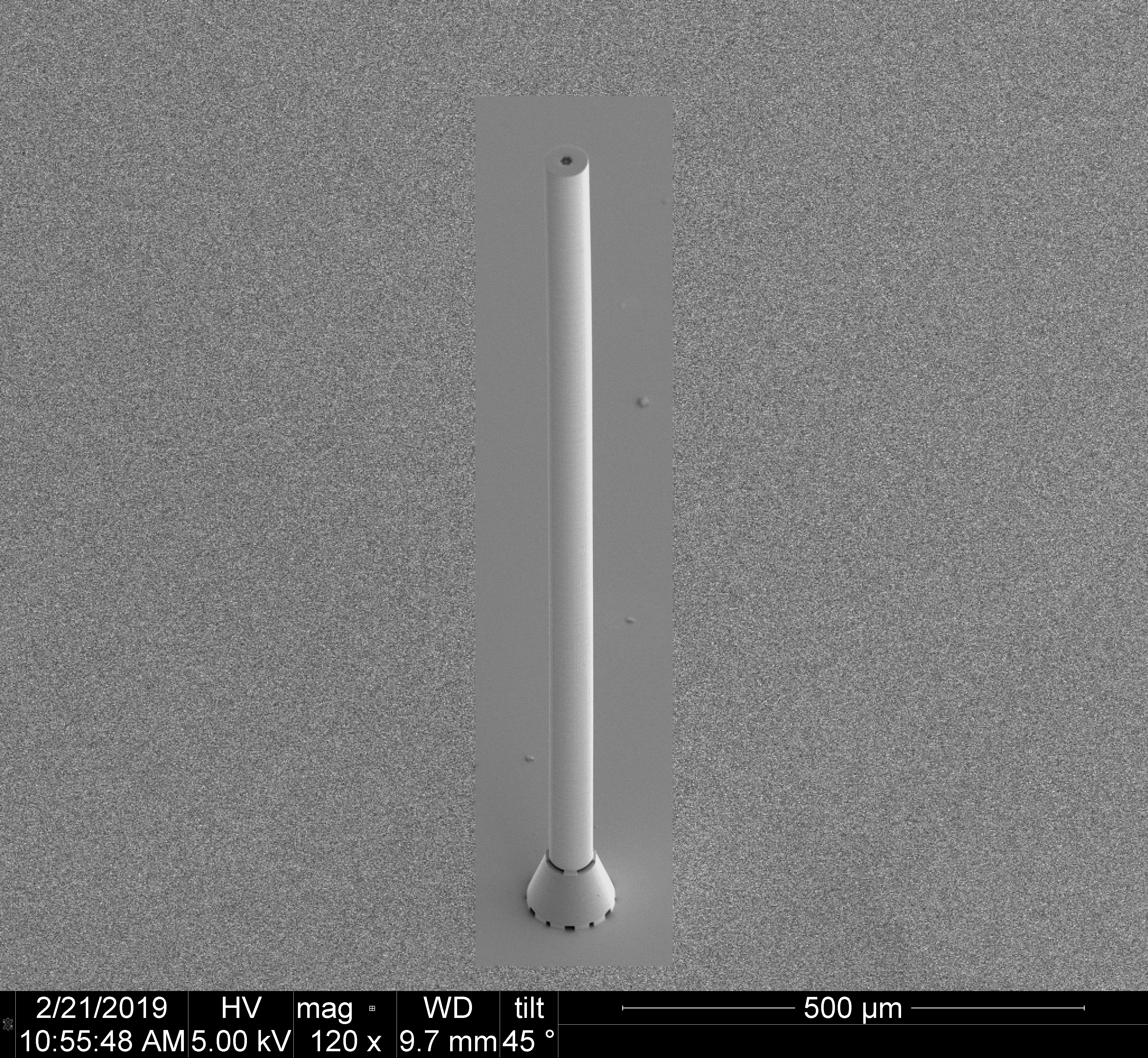}
	\caption{Full length scanning electron micrograph of the 1~mm-long 3D printed HNL \PCF-like segment reported in Fig. \ref{fig:carrellata}a }
	\label{suppl:1mm}
\end{figure}

\pagebreak

\subsection{Estimation of propagation losses}
\label{losses_supp}

\begin{figure}[H] 
	\centering
	\includegraphics[width=0.7\linewidth]{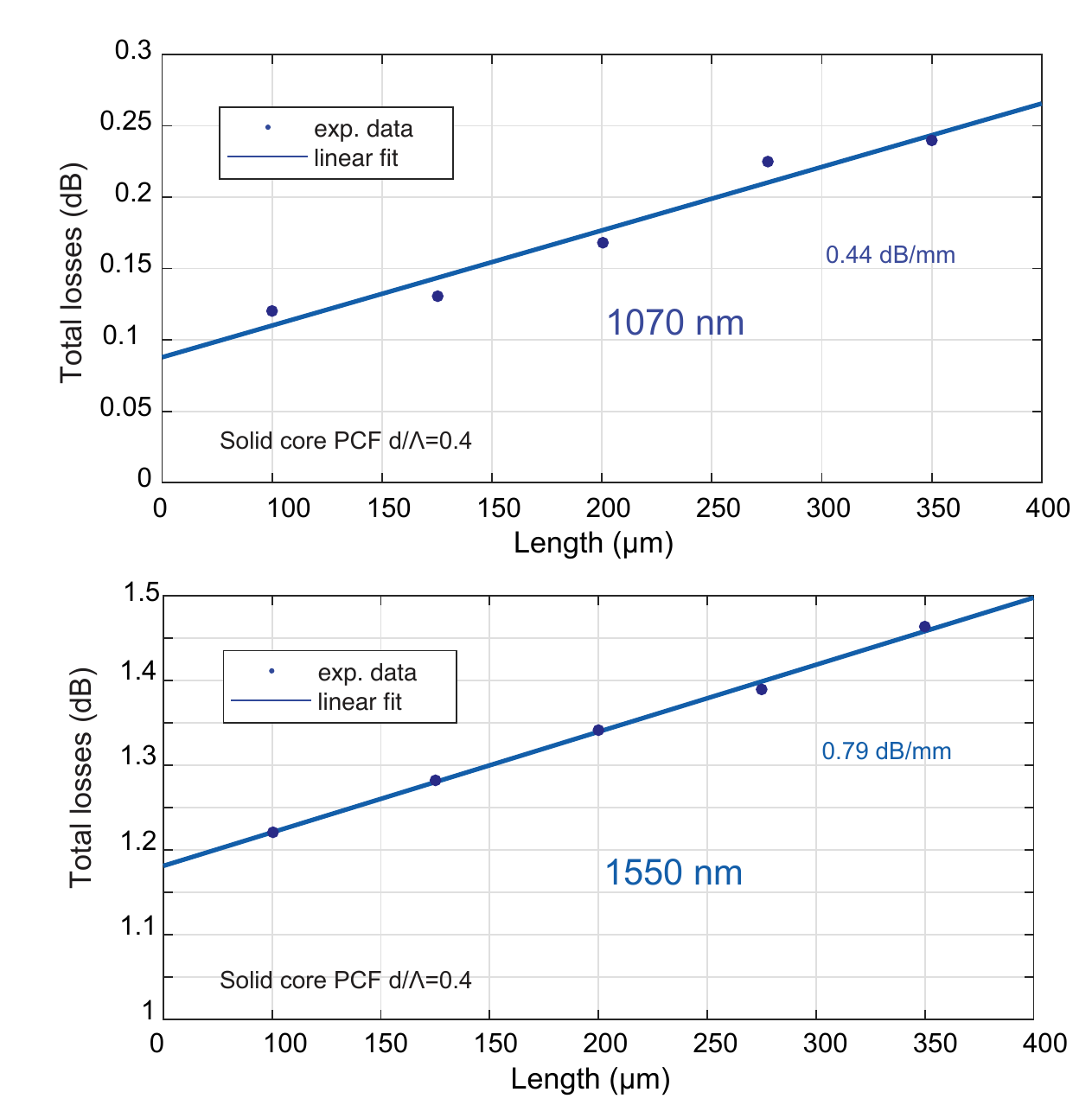}
	\caption{\textbf{Estimation of propagation losses.} We estimated the propagation losses by 3D printing solid-core PCF-like segments with different lengths. We measured the insertion losses as a function of waveguide length, and obtained a propagation loss value by fitting the experimental data. The estimated propagation losses were 0.44~dB/mm ($\pm$ 0.06) at 1070~nm and 0.79~dB/mm ($\pm$ 0.03) at 1550~nm. For comparison, the propagation losses for the used material (IP-Dip) in bulk structures was very accurately measured by Schmid et al. \cite{schmid2019optical} and estimated to be 0.43~dB/mm at 1070~nm and 0.78~dB/mm at 1550~nm. Therefore propagation losses in our 3D printed PCF-like segments are dominated by the the extinction coefficient of the photopolymer.
	} 
	\label{fig:supp_losses}
\end{figure}

\pagebreak

\subsection{Simulation of down-taper segment}
\label{taper_supp}

\begin{figure}[H] 
	\centering
	\includegraphics[width=0.7\linewidth]{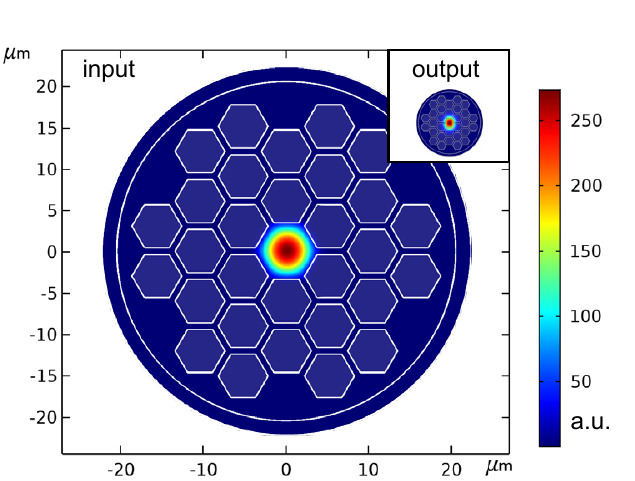}
	\caption{
		\textbf{Design of the down-taper segment.} 
		We show the input and output cross-sections of the down-taper segment, along with the fundamental modes at 1550~nm at each section, as calculated with modal analysis (COMSOL).
		The figures are at the same scale.} 
	\label{fig:taper_sim}
\end{figure}

\subsection{Design and characterization of the fan-out section}
\label{fan-out}

\begin{figure}[]
	\centering
	\includegraphics[width=0.7\linewidth]{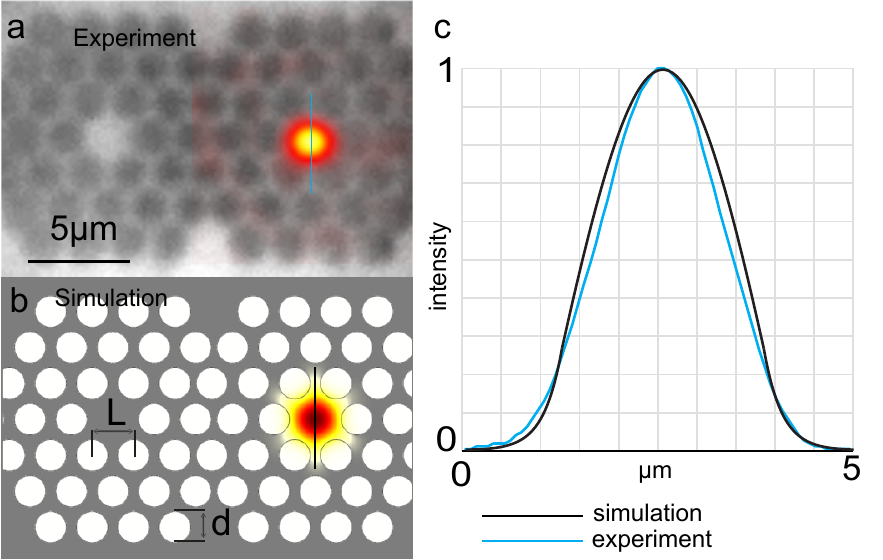}
	\caption{\textbf{Simulation and characterization of fan-out output mode at 1550~nm.}
		a) Optical image. b) Simulation. c) Intensity profiles of experiment and simulation.  (d=1.5 $\mu$m L=2 $\mu$m) }
	\label{fig:fan_out_sim}
\end{figure}

We simulated with modal analysis (COMSOL) the fundamental mode of the output section of the PCF PBS fan-out segment and we compared it with optical images of the mode output from the fabricated structure at 1550~nm (Figure \ref{fig:fan_out_sim}). The mode field diameters are very closely matching between the simulation and the experiment (simulation= 2.92~$\mu$m, experiment= 2.98~$\mu$m)

\pagebreak

\subsection{Coupling length simulations compared with experiments}
\label{Suppl:CLsim_notworking}

\renewcommand{\thefigure}{S.\arabic{subsection}.\arabic{figure}}
\setcounter{figure}{0}

\renewcommand{\thetable}{S.\arabic{subsection}.\arabic{table}}

Real-world effects intrinsic to the 3D micro-printing method and related materials --- such as the discretised geometry in the 3D printing system and a possible anisotropic shrinkage of the structures during the post-exposure development --- could make the 3D printed \PCF-like segment geometrically deviate slightly from the designed PCF geometry. We assessed this effect by comparing the polarisation splitting performances of a designed \PCF \space geometry obtained numerically --- from modal analysis calculations (COMSOL) --- and experimentally on the fabricated structure --- produced using the same  geometry in creating the CAD model file for the 3D printer.
 
\begin{figure}[htbp]
	\centering
	\includegraphics[width=0.6\linewidth]{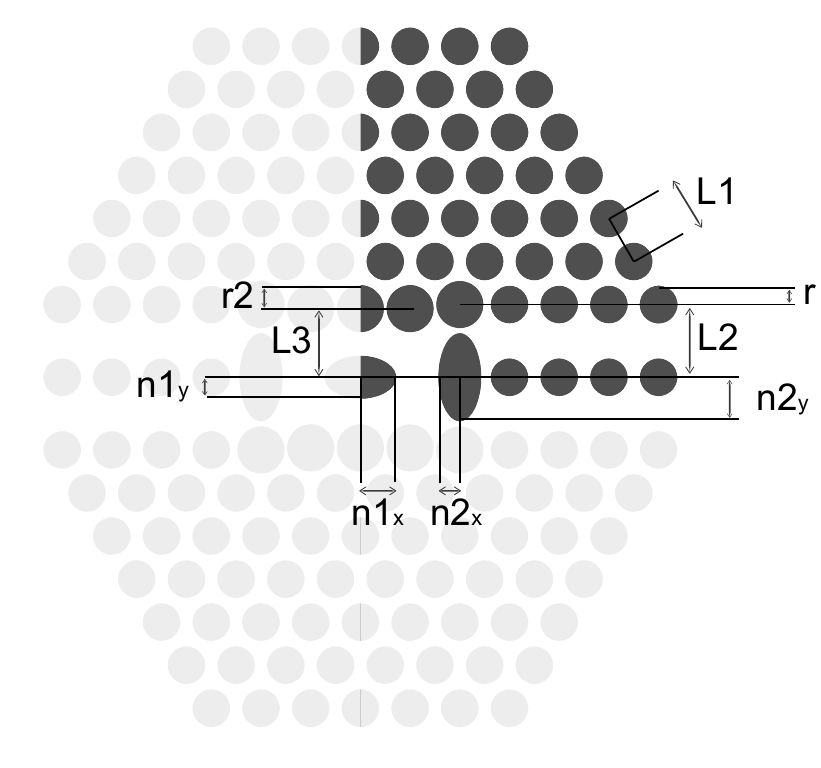}
	\caption{\textbf{
			Geometrical parameters for the dual-core waveguide segment}}
	\label{fig:paralabel}
\end{figure}

We obtained the coupling lengths CLx and CLy, for x and y polarisations respectively, by plugging into Equation \ref{CL_eq} of main text the modal effective indexes calculated for a dual-core \PCF \space with the same geometry proposed by Jiang et al. \cite{Jiang:2014io} (parameters reported in table \ref{tab:Jiang parameters}, with reference to Figure \ref{fig:paralabel}), and using the refractive index of two-photon polymerised IP-Dip material as reported in \cite{schmid2019optical} (1.532 at 1550~nm). 

\begin{table}[h]
	\centering
	\begin{tabular}{|l|l|l|l|l|l|l|l|l|}
		\hline
		L1   & L2   & L3   & r    & r2  & n1\_x & n1\_y & n2\_x & n2\_y \\ \hline
		1.1 & 1.7 & 1.7 & 0.5 & 0.5 & 0.7   & 0.5  & 0.5  & 0.85   \\ \hline
	\end{tabular}
	\caption{Original geometrical parameters of the dual-core waveguide design (dimensions in microns) from \cite{Jiang:2014io}. See figure \ref{fig:paralabel} for  definition of the parameters.}
	\label{tab:Jiang parameters}
\end{table}

\begin{figure}[htbp]
	\centering
	\includegraphics[width=1\linewidth]{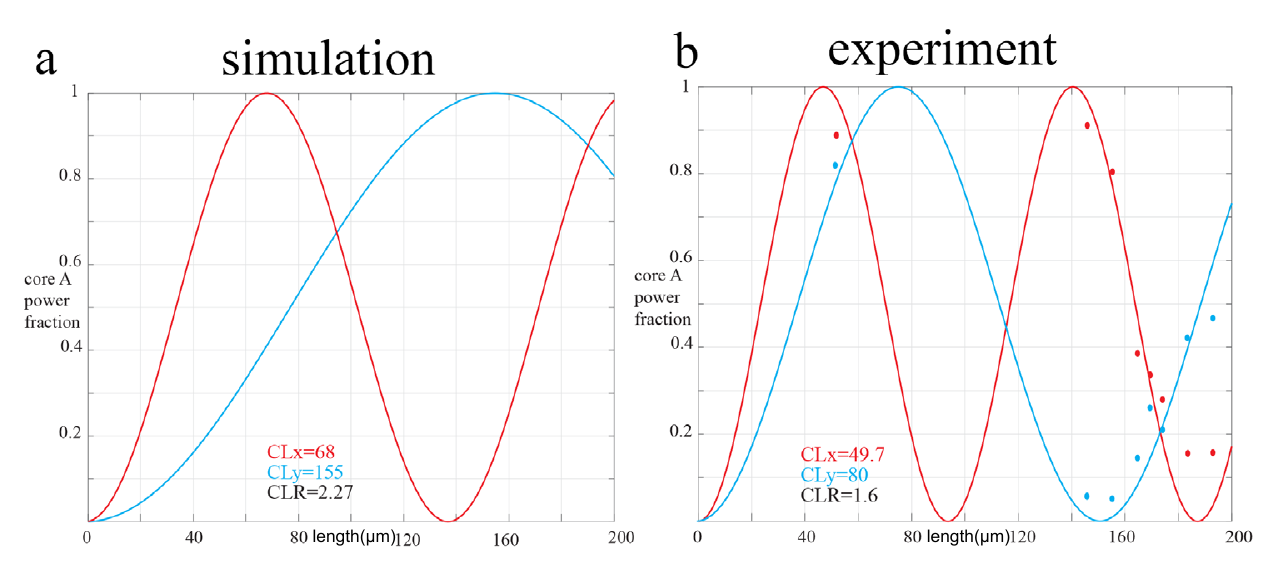}
	\caption{\textbf{Simulations do not match experiments}
		a) Simulation of the power distribution among the cores for different lengths of the dual-core \PCF \space segment for the starting design of our optimisation process.
		b) Experimental values and fitting of the coupling lengths on the same \PCF \space geometry as 3D printed and integrated with the down-taper and fan-out sections}
	\label{fig:CLtrends_suppl_nomatch}
\end{figure}

\begin{figure}[htbp]
	\centering
	\includegraphics[width=0.6\linewidth]{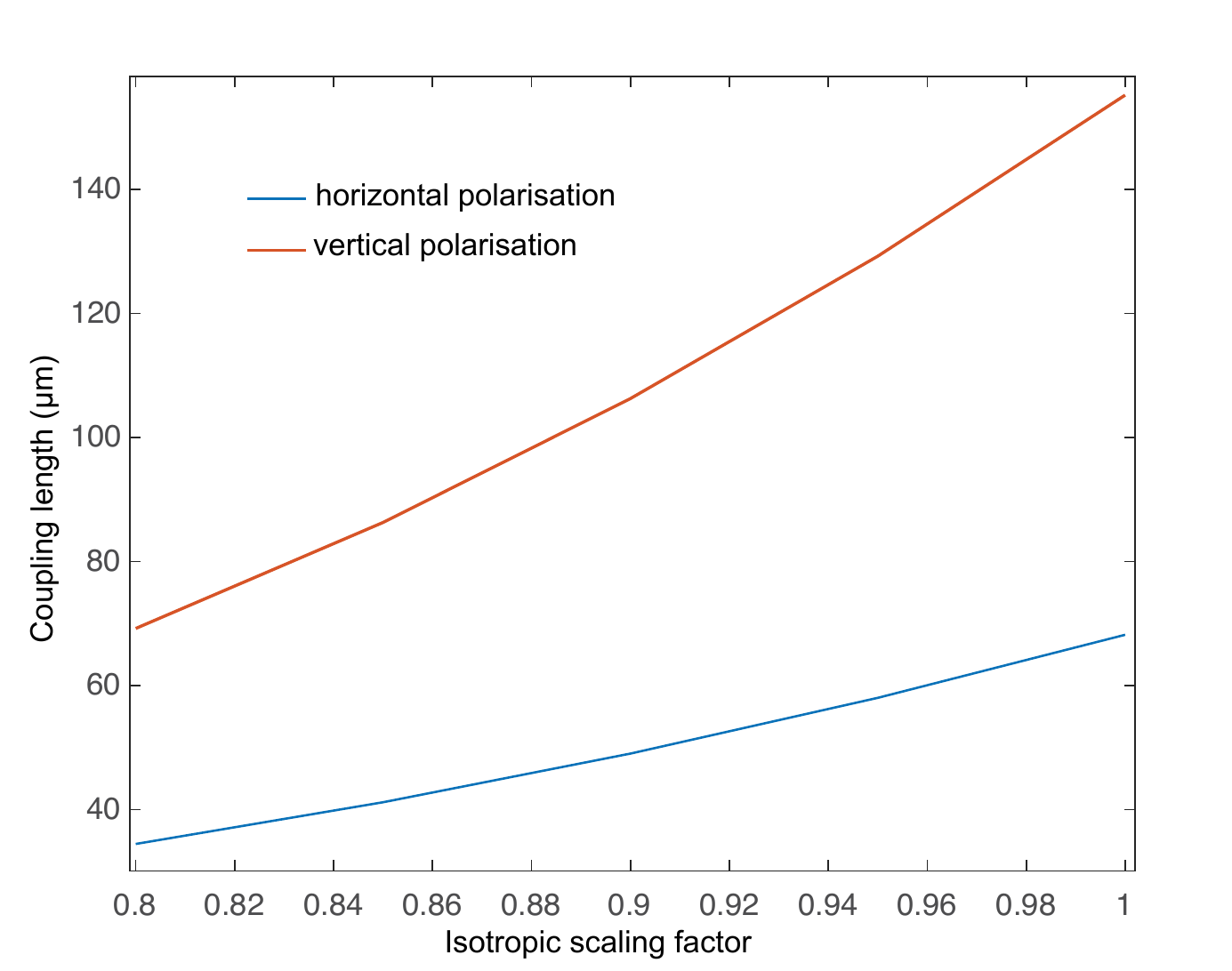}
	\caption{\textbf{Simulations of the effect of isotropic shrinkage on the coupling lengths of the dual-core directional coupler.}
		We simulated the effect of an isotropic scaling of the dual-core section geometry on the coupling lengths at 1550~nm. The shrinkage leads to a significant reduction of the coupling lengths, in line with the observed difference between simulations and experiments (see Figure \ref{fig:CLtrends_suppl_nomatch}).}
	\label{fig:CLtrends_suppl_scaling}
\end{figure}

The coupling lengths for the two polarisations were numerically obtained as
CLx = 155~$\mu$m and CLy = 68.58~$\mu$m, giving a coupling length ratio (CLR) of 2.27 for the simulated structure (Figure \ref{fig:CLtrends_suppl_nomatch}a). 
Although the CLR was not exactly equal to 2, an extinction ratio of 13.5 dB could nonetheless potentially be reached for a 140~$\mu$m length of the structure at the 1550~nm wavelength. 
To obtain the experimental data, we 3D printed a series of structures with the same geometry for the dual-core section and we estimated the coupling lengths at 1550~nm by fitting the power ratio curves (see Figure \ref{fig:CLtrends_suppl_nomatch}b and section S.6). We obtained a CLx = 80~$\mu$m, a CLy = 49.73~$\mu$m, giving a CLR = 1.6, that is significantly different from our numerical calculations. 
The most likely cause of this difference is the slight shrinkage that affects 3D printed structures with the used photopolymer (IP-Dip). Indeed, a critical geometrical parameter affecting the CL for both polarisations is the spacing between the two cores. A slight reduction of this distance, caused by shrinkage, significantly increases the coupling strength of the two cores (see also \cite{Jiang:2014io}) and causes a significant decrease of the CL for both polarisations (see simulated coupling lengths vs. isotropic shrinkage in Figure \ref{fig:CLtrends_suppl_scaling}). 

\pagebreak

\subsection{Estimation of the CLR in 3D printed structures}
\label{suppl:CLR_calc}

\renewcommand{\thefigure}{S.\arabic{figure}}
\setcounter{figure}{5}

To calculate the coupling length ratios of the 3D printed structures during the optimisation process, we collected beam profile images at different wavelengths, as described in \nameref{Methods}. 
The input core was core A and the total output power was $P=P_{core_a}+P_{core_b}$, where $P_{core_a}$ and $P_{core_b}$ are the integrated image intensities on the area of core A and B.

\begin{figure}[htbp]
	\centering
	\includegraphics[width=1\linewidth]{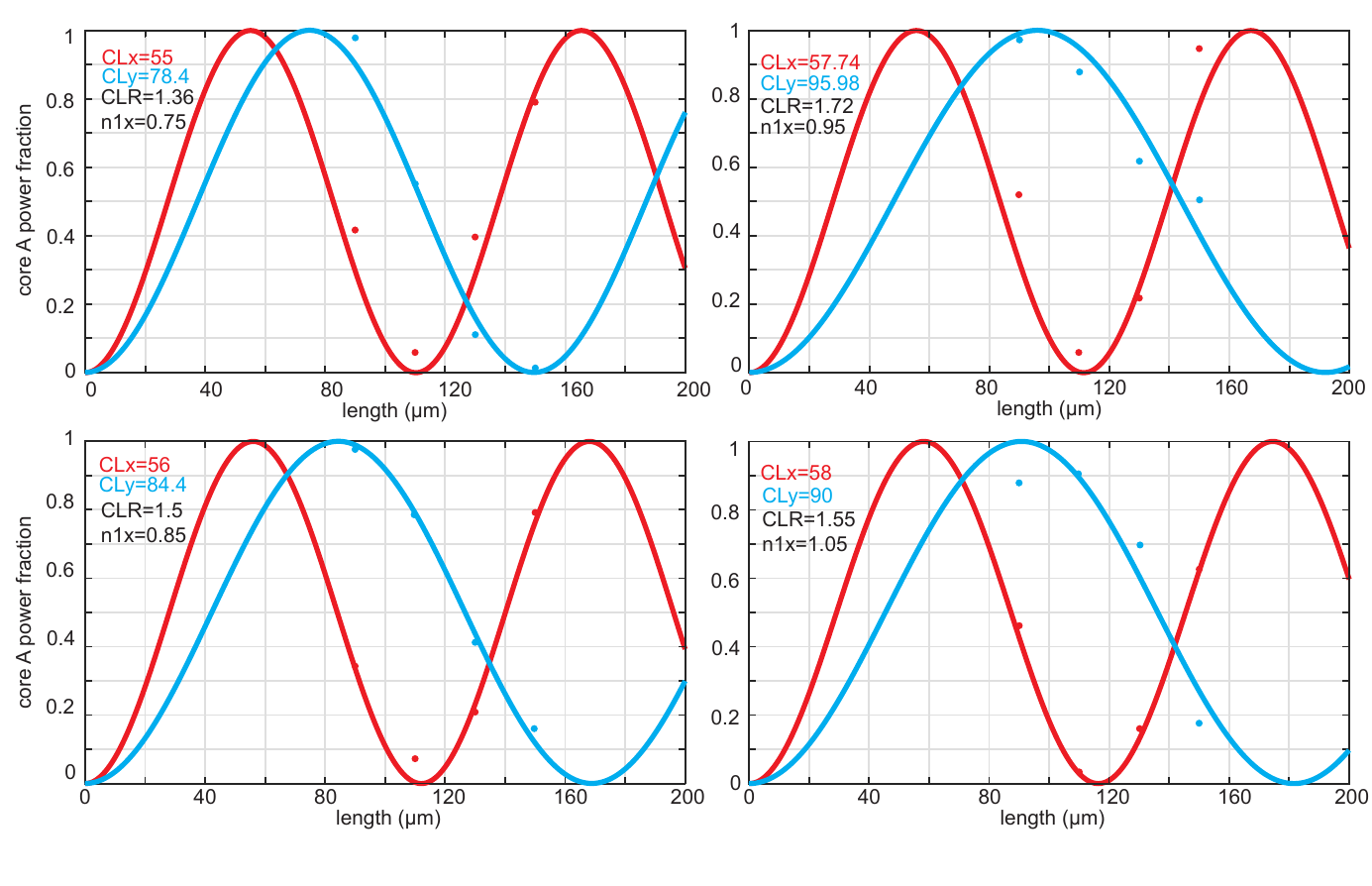}
	\caption{\textbf{Method for the calculation of the coupling lengths}. These plots show the data obtained for the optimisation sample shown in Fig. \ref{fig:optimization_example}, where we investigated four different values of the central hole ellipticity (parameter n1x in Figure \ref{fig:paralabel}) to understand how ellipticity affects the CL. For each value of n1x, we printed four different heights and we calculated the fraction of power on core A for the x and y polarisation, represented as red and blue points on the graphs, respectively.
		Then, to obtain the coupling lengths, we interpolated the data points with a sinusoidal function describing the typical behaviour of directional couplers  \cite{Jiang:2014io}} 
	\label{fig:CLtrends_suppl}
\end{figure}

The fraction of power on core A was calculated as $PF_i=P_{core_b}/P$, where $i$ is the x and y polarisation, represented as red and blue data-points in Fig. \ref{fig:CLtrends_suppl}. 
These datapoints have been interpolated with a function $PF_i=sin^2((pi*length)/(2*CL_i))$ to get the $CL_i$, which are the coupling lengths for the x and y polarisation.

\pagebreak

\subsection{Optimisation procedure for the \PCF \space PBS}
\label{Suppl:optimization}

\setcounter{table}{0}

The parameters that we optimised to reach a CLR of 2, summarised in Fig. \ref{fig:OptimizationProcess}, were the position of the 6 round holes close to the cores, the ellipticity of the central hole separating the two cores, and the diameters of the 10 round holes closest to the cores.
With reference to Figure \ref{fig:paralabel}a, the initial parameters of our dual-core waveguide (Fig. \ref{fig:OptimizationProcess}, "Starting design") are reported in Table \ref{tab:initial parameters}.

\begin{table}[h]
	\centering
	\begin{tabular}{|l|l|l|l|l|l|l|l|l|}
		\hline
		L1   & L2   & L3   & r    & r2  & n1\_x & n1\_y & n2\_x & n2\_y \\ \hline
		1.21 & 1.77 & 1.77 & 0.48 & 0.48 & 0.9   & 0.55  & 0.55  & 1.1   \\ \hline
	\end{tabular}
	\caption{Initial parameters of the dual-core waveguide (dimensions in microns). See figure \ref{fig:paralabel} for definition of the parameters.}
	\label{tab:initial parameters}
\end{table}

\begin{figure}[htbp]
	\centering
	\includegraphics[width=1\linewidth]{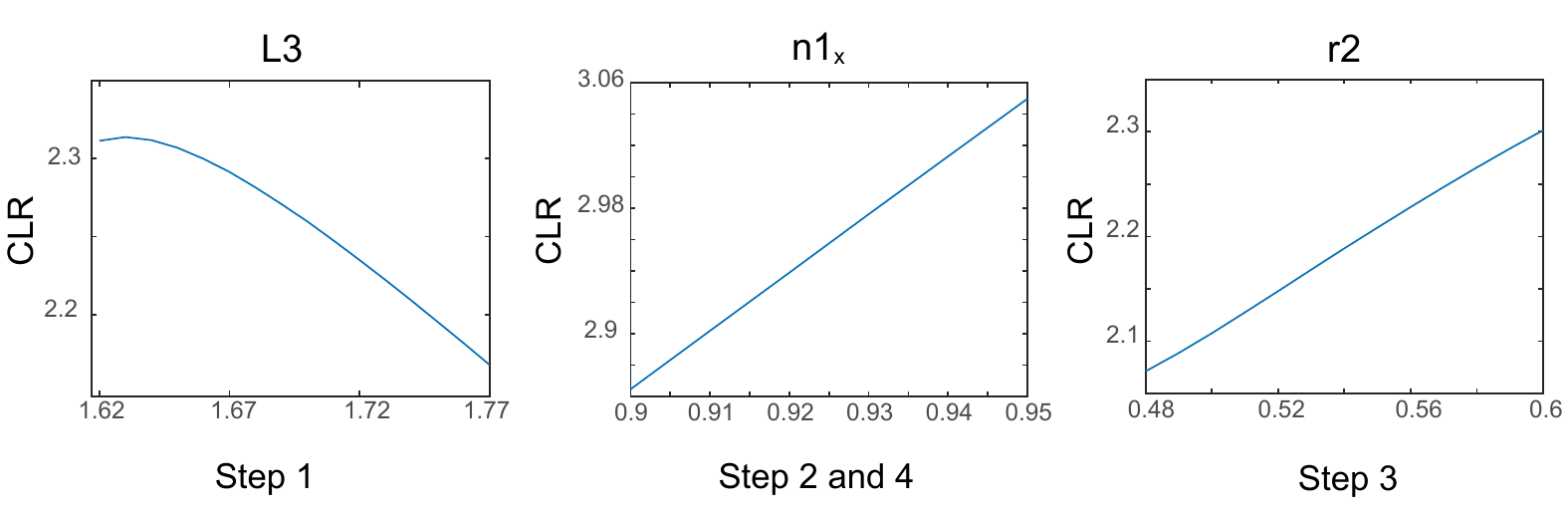}
	\caption{\textbf{Guidelines from simulations.} Simulated CLR vs. the variation of selected geometrical parameters.}
	\label{fig:CLtrends_simulation}
\end{figure}

As shown in Figure \ref{fig:OptimizationProcess}a, in Step 1 we changed the parameter L2 from 1.77 $\mu$m to 1.67 $\mu$m. In Step 2 we changed n1$_x$ from 0.9 to 0.95. In Step 3 we changed r2 from 0.48 to 0.6 and in Step 4 we changed n1$_x$ from 0.95 to 0.9. The choice of changing these particular geometrical parameters has been driven by the simulation of the 
trend in the variation of the CLR associated with the change of those parameters (Figure \ref{fig:CLtrends_simulation}).

\pagebreak

\subsection{Characterisation of the \PCF \space PBS structures with confocal fluorescence microscopy}
\label{suppl:confoca}

\begin{figure}[htbp]
	\centering
	\includegraphics[width=1\linewidth]{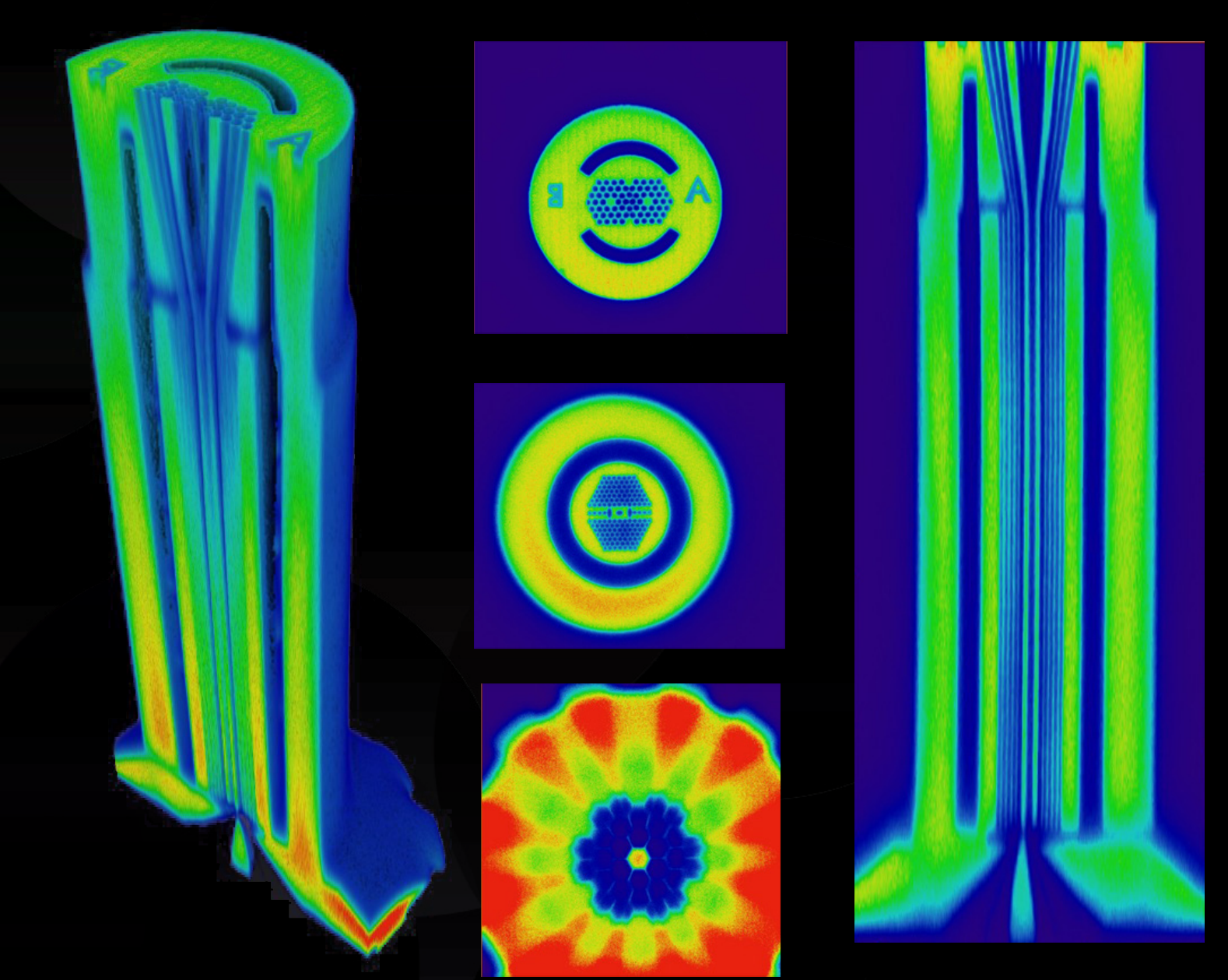}
	\caption{Rendered three dimensional stack and cross-sections obtained with confocal fluorescence microscopy of one complete \PCF \space PBS structure}
	\label{fig:confocal}
\end{figure}

We analysed one complete 3D printed \PCF  \space PBS structure with fluorescence confocal microscopy to ascertain that the longitudinal channels of the \PCF-like segments were properly and uniformly developed and confirm that the inner parts of the structure were printed as designed.
This method takes advantage of the intrinsic weak fluorescence of polymerised material due to the presence of photoinitiator molecules; thus, a fluorescence signal --- excited at 488nm and detected in the 500-600 nm band --- is generated, even without introducing external contrast agents. To achieve high-resolution images across the whole length of the 3D-printed structure, we used an immersion silicon oil (Cargille, Laser Liquid Code 1074) with a finely tuned refractive index that matches the refractive index of the polymerised photoresist at the excitation wavelength.
The 3D stack (Fig. \ref{fig:confocal}) obtained with a Zeiss LSM880 confocal microscope clearly shows a uniform structure along its length, with accurate hole pattern fabrication and well-developed channels.

\end{document}